\documentclass[sigconf]{acmart}
\PassOptionsToPackage{table}{xcolor}
\usepackage[table]{xcolor}
\usepackage{soul}
\usepackage{verbatim}
\usepackage{listings}
\usepackage{colortbl}

\definecolor{codegreen}{rgb}{0,0.6,0} 

\lstdefinestyle{mystyle}{
    backgroundcolor=\color{white}, 
    commentstyle=\color{codegreen}, 
    keywordstyle=\color{blue}, 
    stringstyle=\color{codegreen}, 
    basicstyle=\ttfamily\footnotesize, 
    breaklines=true, 
    captionpos=b, 
    numbers=left, 
    keepspaces=true }
\newcommand{\rev}[1]{{#1}} 
\newcommand{\revt}[1]{{#1}} 

\AtBeginDocument{%
  }

\copyrightyear{2025}
\acmYear{2025}
\acmConference[CHI '25]{CHI Conference on Human Factors in Computing
Systems}{April 26-May 1, 2025}{Yokohama, Japan}
\acmBooktitle{CHI Conference on Human Factors in Computing Systems (CHI
'25), April 26-May 1, 2025, Yokohama,
Japan}
\acmDOI{10.1145/3706598.3714222}
\acmISBN{979-8-4007-1394-1/25/04}

\begin{document}

\title{Conversation Progress Guide : UI System for Enhancing Self-Efficacy in Conversational AI}

\author{Daeun Jeong}
\affiliation{%
  \institution{Jeonbuk National University}
  \city{Jeonju}
  \country{Republic of Korea}
}
\email{daeun3736@jbnu.ac.kr}
\orcid{XXXX}

\author{Sungbok Shin}
\authornote{The work was done while the author was affiliated with the University of Maryland, College Park.}
\affiliation{%
  \institution{Inria, Universit\'e Paris-Saclay}
  \city{Saclay}
  \country{France}
}
\email{sungbok.shin@inria.fr}
\orcid{0000-0001-6777-8843}

\author{Jongwook Jeong}
\authornote{corresponding author.}
\affiliation{%
  \institution{Jeonbuk National University}
  \city{Jeonju}
  \country{Republic of Korea}
}
\email{jwjeong55@jbnu.ac.kr}

\renewcommand{\shortauthors}{Jeong et al.}

\begin{abstract}
In this study, we introduce the Conversation Progress Guide (CPG), 
a system designed for text-based conversational AI interactions that provides a visual interface to represent progress.
Users often encounter failures when interacting with conversational AI, which can negatively affect their self-efficacy---an individual's belief in their capabilities, reducing their willingness to engage with these services. 
The CPG offers visual feedback on task progress, providing users with mastery experiences, a key source of self-efficacy. 
To evaluate the system's effectiveness, we conducted a user study assessing how the integration of the CPG influences user engagement and self-efficacy. 
Results demonstrate that users interacting with a conversational AI enhanced by the CPG showed significant improvements in self-efficacy measures compared to those using a conventional conversational AI.

\end{abstract}




\maketitle

\section{Introduction}
\label{sec:01-introduction}
Self-efficacy, or belief in one's abilities, is crucial for learning and work efficiency. 
If generative conversational AI lowers self-efficacy, users may develop negative views and avoid these services. 
Systems like ChatGPT and Copilot have gained widespread use, but their inherent limitations lead to issues such as misunderstanding user intent, abrupt conversation endings, repetitive responses, and hallucinations, resulting in user dissatisfaction.

These issues are particularly important to address because in tasks that require multi-turn conversations, they can have negative impacts on users' self-efficacy beyond mere dissatisfaction. 
Doubts about information accuracy and transparency, as well as hallucinations, can impose excessive cognitive load on users performing tasks, potentially leading to a decrease in self-efficacy~\cite{baumeister2018ego,feldon2023direct}. 
Furthermore, when repetitive responses and failures to understand user intent occur, users experience failures that become major factors in lowering self-efficacy.

It is well known that self-efficacy greatly affects task performance and learning efficiency~\cite{bandura1982microanalysis,pajares1994role}, and interfaces designed to enhance self-efficacy already exist~\cite{rodrigues2021promoting,evans2021designing,al2022development}. However, these are mostly limited to interfaces that improve specific learning situations using digital devices. While applications utilizing conversational agents to increase self-efficacy have been proposed, research aiming to promote self-efficacy in the context of using generative conversational AI has not yet been conducted. As the importance and application domains of generative conversational AI continue to expand, improving self-efficacy in this environment is a significant challenge that needs to be addressed.

In this paper, we present the Conversation Progress Guide~(CPG), a visual interface that illustrates the progress of a conversation, contributing to the enhancement of users' self-efficacy. Our research goal is not to improve the quality or performance of existing conversational services but to achieve positive effects on self-efficacy through an interface designed to enhance self-efficacy without altering the underlying technology. Motivated by the effect of progress bars~\cite{mayer2002multimedia}, we designed a progress interface for conversational AI as a method to help strengthen users' self-efficacy.

\textbf{(1) The CPG indicates the completion of the subtasks that comprise the user's task.} The problem-solving process using conversational AI often includes many interactions, and this process may prevent users from gaining successful experiences, potentially causing frustration. By presenting the completion of subtasks to the user, the system offers mastery experiences, the most powerful source of self-efficacy.

\textbf{(2) The CPG provides users with feedback on task progress by displaying a visual progress bar.} While conversational AI can present feedback or encouragement through dialogue, it does not provide fixed performance feedback visually. Performance feedback on current progress aids self-efficacy and encourages users to continue the task~\cite{azevedo2004does,deci2000and}. Additionally, the progress bar can positively affect users' physiological and emotional states by presenting the task's progress status to reduce anxiety.

To verify the effectiveness of the CPG, we developed a prototype and conducted a user study. 
Participants were divided into a control group using the default conversational AI interface and an experimental group using the AI with the CPG applied. 
Both groups performed the same task. 
To evaluate the impact of the CPG on users, we measured self-efficacy before and after task performance, cognitive load, satisfaction, task time, and interaction count with the conversational AI.

Analysis revealed that both groups improved in self-efficacy, with significantly greater gains in the CPG group. 
This shows that CPG fosters self-efficacy, encouraging more proactive use of conversational AI in problem-solving.
Furthermore, no significant differences in other measures were found, confirming the system does not increase cognitive load.


In summary, our work makes the following contributions. 
We have:
\begin{enumerate}
    \item introduced the CPG, a novel interface concept that illustrates task progress in conversational AI to enhance user self-efficacy.
    \item developed a structured design approach for the CPG, which evaluates the completion of subtasks and provides users with structured feedback to maintain engagement during interactions.
    \item conducted a user study demonstrating that the CPG enhances self-efficacy while maintaining cognitive load, task performance, and user satisfaction.
\end{enumerate}

\section{Background and Related Work}
\label{sec:02-background}

We present the background about self-efficacy, interfaces that enhance self-efficacy, and failures on generative conversational AI. 

\paragraph{Self-Efficacy} Self-efficacy refers to an individual's belief in their ability to accomplish specific goals~\cite{bandura1978self,bandura1986social,bandura1997self}. 
It reflects confidence in one's ability to control motivation, behavior, and environmental factors. 
The level of self-efficacy influences the goals individuals set, the energy they invest in achieving those goals, and their behavioral performance. 
Low self-efficacy diminishes motivation and leads individuals to avoid challenges~\cite{bandura1997self}. 
It also increases anxiety, reduces persistence, and ultimately results in decreased learning ability and performance~\cite{schwarzer1995measures,pajares1996self}. In his theory of self-efficacy, Bandura identifies four sources that build and enhance self-efficacy: mastery experiences, vicarious experiences, social persuasion, and physiological and emotional states. Mastery Experiences are direct personal achievements that most effectively enhance self-efficacy. Vicarious Experiences are the experiences gained by observing others' success, which help individuals believe they can succeed as well. Social Persuasion is the positive encouragement or feedback from others that promotes self-efficacy. Physiological and Emotional States are the physical and emotional conditions that influence confidence, with positive states enhancing self-efficacy.

\paragraph{Interfaces that Enhance Self-Efficacy} Park et al. developed a conversational agent-based application designed to enhance the self-efficacy of children with Attention Deficit Hyperactivity Disorder (ADHD)~\cite{park2023utilizing}. Low cognitive ability children with ADHD often have fewer successful experiences and consequently lower self-efficacy compared to their peers. 
Through interactions with the conversational agent, children are able to plan, execute, and review their task performance. The application encourages consistent task completion, which was found to result in an improvement of self-efficacy among the participants.

Huang et al. proposed strategies to build the four sources of self-efficacy within an online learning environment~\cite{huang2020better}. Each of the four strategies demonstrated a positive effect on improving self-efficacy, and applying all four strategies together proved more effective. 
Among their strategies, Attributional Feedback entailed displaying effort-based feedback after each practice problem, with messages tailored to whether participants solved the problem correctly or incorrectly, reducing anxiety and improving self-efficacy.

Our approach aligns with the mechanisms used in these previous studies to strengthen self-efficacy. However, limited research has explored self-efficacy in the context of conversational AI. Given that conversational AI may negatively impact self-efficacy, further investigation is needed.

\begin{table*}[ht]
\centering
\caption{Categories of Dissatisfaction in Conversational AI Responses}
\label{tab:categories}
\begin{tabular}{lll}
\toprule
\textbf{Category} & \textbf{Description} & \textbf{References} \\ 
\midrule
Intent Understanding & Failing to accurately reflect user intent or context & \cite{kim2024understanding,lin2024interpretable} \\ 
\rowcolor{gray!18} Content Depth and Originality $\quad$ & Responses are overly generic or lack originality & \cite{kim2024understanding,lin2024interpretable} \\ 
Information Accuracy & Providing incorrect or inconsistent information & \cite{kim2024understanding,lin2024interpretable,see2021understanding} \\ 
\rowcolor{gray!18}
Transparency & Lack of clarity on the basis or rationale for responses & \cite{kim2024understanding} \\ 
Refusal to Answer & Avoiding or refusing to answer specific questions & \cite{see2021understanding,kim2024understanding} \\ 
\rowcolor{gray!18}
Content Ethics and Integrity & Including unethical or harmful content & \cite{see2021understanding,kim2024understanding}  \\ 
Response Format and Attitude & Inappropriate format, tone, or length of responses & \cite{kim2024understanding,lin2024interpretable,see2021understanding} \\
\rowcolor{gray!18}
Bot Repetitive & Repeating the same information excessively & \cite{see2021understanding} \\ 
Bot Hallucination & Generating non-existent or incorrect information & \cite{see2021understanding} \\
\rowcolor{gray!18}
Sudden End & Abruptly ending the conversation without resolution $\quad$& \cite{lin2024interpretable} \\ 
\bottomrule
\end{tabular}
\end{table*}

\paragraph{Failures on Generative Conversational AI}
See and Manning~\cite{see2021understanding} conducted an earlier case study on the neural generative model deployed in the social chatbot, Chirpy~\cite{paranjape2020neural}. 
They investigated conversations between users and the bot, analyzing the relationship between errors and user dissatisfaction. Based on their findings, they defined taxonomies to systematically categorize these issues.
Kim et al.~\cite{kim2024understanding} conducted a study to understand the dissatisfaction users experience during interactions with generative conversational AI. 
They reviewed 59 papers to collect failure cases and performed open coding on these instances, ultimately classifying user dissatisfaction factors arising from generative conversational AI responses and interactions into seven categories. 
Similarly, Lin et al.~\cite{lin2024interpretable} extracted signals of satisfaction from user inputs when using generative conversational AI to identify patterns of satisfied and dissatisfied conversations, summarizing the related dissatisfaction patterns in the process.
By synthesizing the categorized dissatisfaction factors from these three studies, we present them into ten categories (see Table~\ref{tab:categories}).

These studies indicate that users experience a wide range of dissatisfaction when interacting with generative conversational AI. 
However, users rarely adopt strategies to address these issues, and even when they do, 72\% of the dissatisfactions remain unresolved~\cite{kim2024understanding}. 
These dissatisfaction and failure experiences may result in a decrease in users' self-efficacy when using generative conversational AI. 
To address this issue, our approach focuses on utilizing the partial mastery experiences that users acquire. 
We aim to enhance users' self-efficacy by visually displaying the achievement of subtasks and associated progress during their interactions with the AI.

\section{Solution Approach}
Below we describe our solution approach. 

\begin{figure*}[h]
  \centering
  \includegraphics[width=0.8\linewidth]{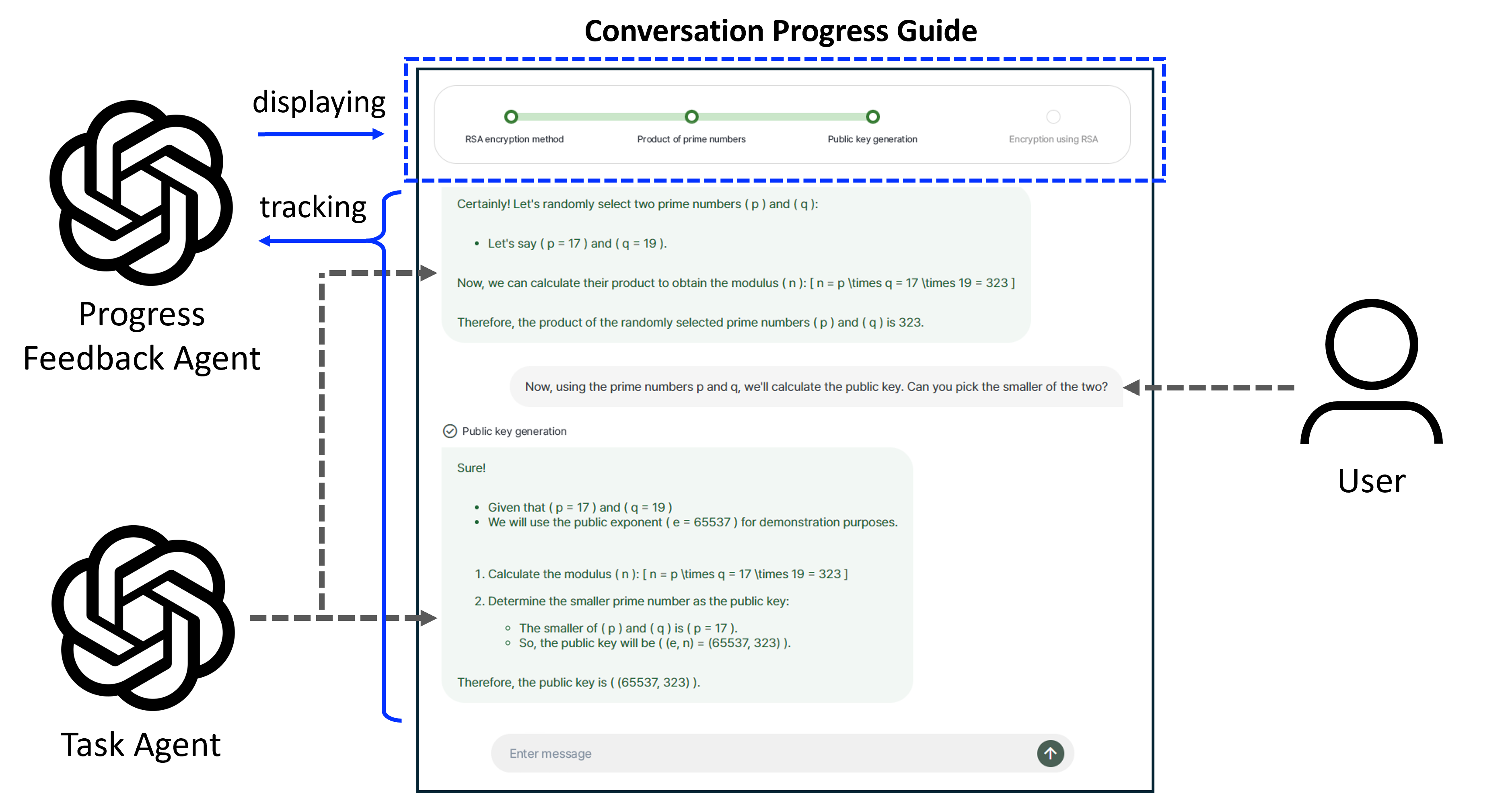}
  \caption{The system architecture of the CPG, highlighting the interaction between the Task Agent and the Progress Feedback Agent in providing feedback to users.}
  \Description{This figure depicts the architecture of the Conversation Progress Guide, showing the interaction between the user, Task Agent, and Progress Feedback Agent. The Task Agent responds to user queries, while the Progress Feedback Agent monitors progress and provides feedback through a visual progress bar.}
  \label{fig:1}
\end{figure*}

\subsection{Design Requirements}
We designed CPG based on three key design requirements. 
First, we prioritized on the efficient use of screen space.
Accordingly, we allocated a significant portion of the interface showing the conversation history between the user and conversational AI, as it is the main part of our work. 
Second, we wanted users to easily understand the progress. 
We focused on the design's affordances to ensure users could easily understand the tool's utility.
Third, we wanted the tool to support continuous progress monitoring and explicitly communicate achievements to reinforce users' self-efficacy. 
Providing clear feedback on subtask completion helps users build a sense of accomplishment, which is vital for maintaining motivation.
Based on these considerations, we decided that a progress bar, in combination with markers is the \textit{desiderata} to support our goal.
Its compact format efficiently conveys information, while markers for subtask completion provide clear feedback~\cite{myers1983incense,myers1985importance,marwan2021promoting}.

\subsection{Conversation Progress Guide Architecture}

CPG requires two AI agents to operate, as shown in Figure~\ref{fig:1}. 
The first agent is the Task Agent, which functions identically to a conventional conversational AI; users interact with this agent to solve problems. 
The second agent is the Progress Feedback Agent, which monitors the task's progress and evaluates the user's advancement.

This agent analyzes the user's inputs and the Task Agent's responses during their interaction to evaluate whether the conversation aligns with the set goal and to determine if the subtasks required to achieve that goal have been completed. 

When the agent evaluates that a subtask has been completed, it provides visual feedback by activating an additional step marker on the progress bar. 
This allows users to visually confirm the completion status of subtasks, their overall progress, and whether the task has been fully accomplished during the conversation. 

\subsection{User Interface}

\begin{figure}[h]
  \centering
  \includegraphics[width=0.45\textwidth]{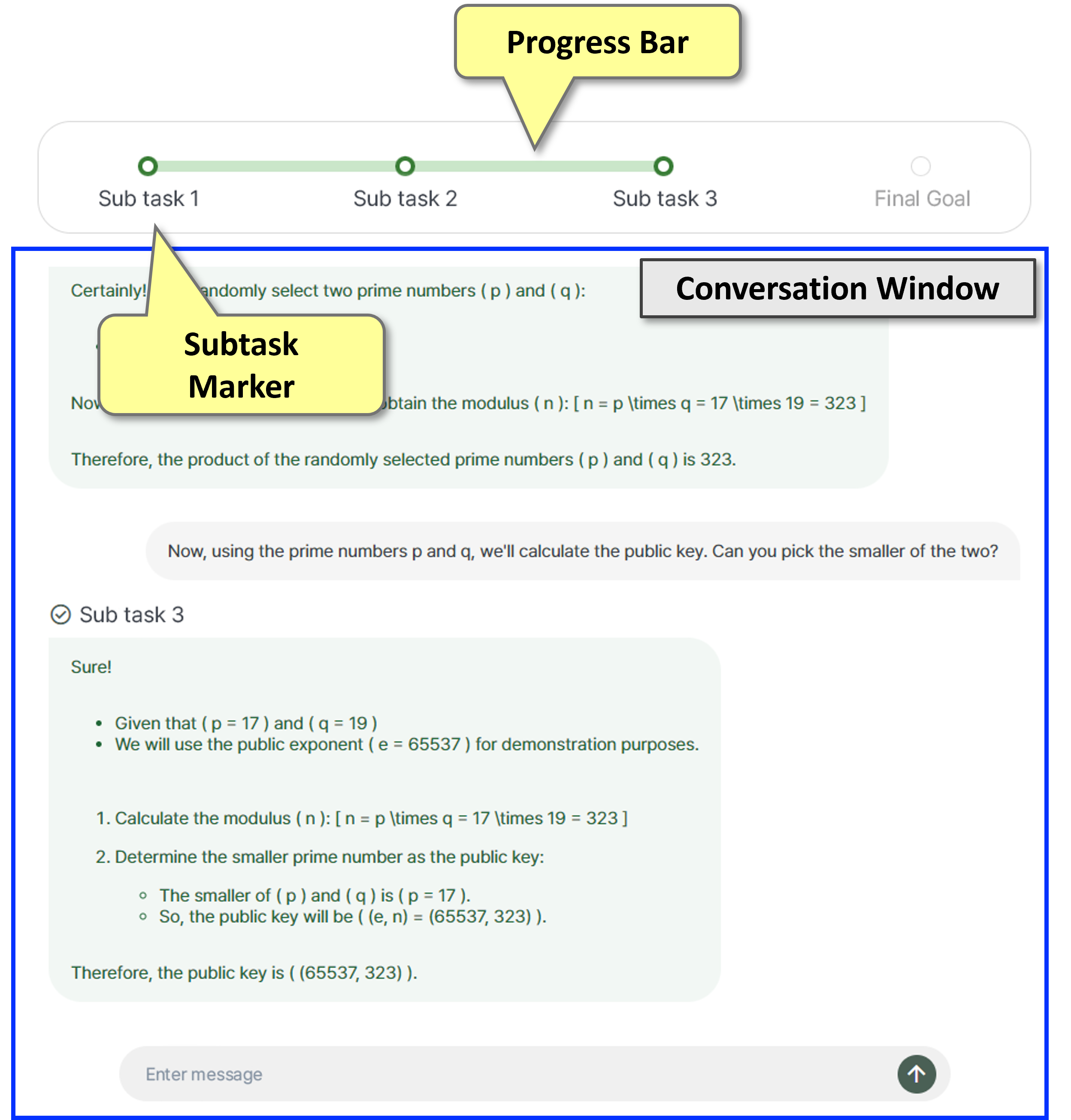}
  \caption{User Interface of Conversation Progress Guide}
  \Description{This figure shows the user interface of the Conversation Progress Guide. It includes a progress bar that updates as users complete subtasks, providing visual feedback on their progress through a series of conversation-based tasks.}
  \label{fig:2}
\end{figure}

The user interface of the CPG comprises two core components: the Progress Bar and the Subtask Marker (see Figure~\ref{fig:2}). 
These elements collaboratively offer a visual representation of task progress and feedback on subtask completion.

\paragraph{\textbf{Progress Bar}}
The CPG offers two main visual elements: the Progress Bar and the Subtask Marker. 
The progress bar is an indicator that visually supports users in recognizing the progress of conversations in problem-solving. 
Initially, an empty progress bar is presented with all markers deactivated; only the marker symbolizing the final goal is displayed in a deactivated state at the end of the progress bar. 
This design helps users clearly recognize the goal they need to achieve.

Through interactions with the Task Agent, users incrementally accomplish predefined subtasks. 
When a subtask is completed, an activated marker is added to the progress bar in the predefined order, displaying only the markers for subtasks the user has achieved to highlight their efforts in the problem-solving process.

This design is based on the concepts of goal setting and motivation presented in social cognitive theory. Visual feedback indicating progress toward goal achievement can act as a factor in promoting self-efficacy and help users maintain motivation toward their goals~\cite{schunk2021self}. Therefore, the progress bar in this study serves as an important tool that helps users clearly recognize their goals and reinforces the belief that they are making progress toward achieving them.

\paragraph{\textbf{Subtask Marker}} 
The Subtask Marker is an indicator that visually supports users in recognizing the accomplishment of subtasks. 
Tasks requiring multi-turn conversations have a high likelihood of involving multiple failures in the process of achieving the goal. 
Mastery experience, one of the main sources of self-efficacy, has the most powerful influence on self-efficacy; negative experiences such as failures or mistakes can significantly diminish it~\cite{gale2021mixed}. 
We focused on enhancing self-efficacy by emphasizing the sense of accomplishment when even small tasks are successfully completed, rather than dwelling on the negative impacts of failures.

To this end, we subdivided the user's goal into subtasks and designed the progress bar into multiple stages. 
Additionally, instead of exposing all subtasks, we adopted a method of adding markers only for successful subtasks, allowing users to actively experience mastery. 
This design aims to encourage users to focus on successful experiences rather than on failed tasks.

Bandura~\cite{bandura1986social} presents three elements that explain active mastery: first, the event must actually occur; second, the individual must directly perform the action and experience a sense of success; third, the event must contribute to achieving important immediate or long-term goals. 
Based on these elements, we designed the system to enhance self-efficacy by allowing users to incrementally experience how their small achievements contribute to accomplishing larger goals~\cite{schwarzer2014self}.

\section{User Study and Results}
This section describes the user study conducted to evaluate the effectiveness of the CPG. We present the user study, and results, including a comparative analysis of the control and experimental groups.

\subsection{User study}
To evaluate the effectiveness of the proposed CPG, we developed a prototype and conducted a user study. 
Participants were recruited and randomly assigned into two groups: a control group using the default conversational AI interface, and an experimental group using the AI with the CPG applied. 
Both groups were instructed to perform the same task using the conversational AI, and we compared their processes and outcomes.

\begin{table*}[ht]
\centering
\caption{RSA Encryption Task Self-Efficacy Survey Questions}
\renewcommand{\arraystretch}{1.3}
\begin{tabular}{ll}
\hline
\toprule
\textbf{No.} & \textbf{Survey Question} \\
\midrule
1 & Do you think you fully understand the basic concepts of RSA encryption? \\ 
\rowcolor{gray!18}
2 & Are you confident that you can obtain the RSA private key using prime numbers $p$ and $q$? \\
3 & Do you think you can correctly calculate the public key using the private keys $p$ and $q$? \\
\rowcolor{gray!18}
4 & Are you confident that you can accurately derive the private key $d$ based on the calculated public key? \\
5 & Do you think you can successfully encrypt the string ``JBNU\_CSAI'' using RSA encryption? \\
\rowcolor{gray!18}
6 & Are you confident that you can clearly explain the entire RSA encryption process? \\
\bottomrule
\end{tabular}
\label{tab:rsa_self_efficacy_survey}
\end{table*}

\begin{table*}[h]
\centering
\caption{Satisfaction Survey Questions.}
\renewcommand{\arraystretch}{1.3}
\begin{tabular}{ll}
\hline
\toprule
\textbf{No.} & \textbf{Survey Question} \\
\midrule
1 & Were you overall satisfied while using this system? \\
\rowcolor{gray!18}
2 & How helpful do you think this system was in performing the task? \\
3 & Would you like to use this system again? \\
\rowcolor{gray!18}
4 & Were you satisfied with the design of this system? \\
\bottomrule
\end{tabular}

\label{tab:satisfaction_survey}
\end{table*}

In total, 22 participants took part in the experiment, comprising 14 females and 8 males. Their average age was 22.05 years, with the youngest being 20 and the oldest 26. 

We recruited participants who were undergraduate students at a university majoring in Computer Science, and were knowledgeable about information encryption tasks. 
These criteria aimed to recruit motivated individuals with an interest in solving the assigned tasks, thereby reflecting a real-world scenario in the experimental results. 
Additionally, this approach minimized biases stemming from variations in participants’ AI familiarity and task expertise. 
Each participant had completed at least three semesters, ensuring a consistent baseline of academic experience. 
Notably, none of the participants had a clear understanding of the steps involved in RSA encryption. Each participant received approximately \$9 as a reward.

Prior to the experiment, we provided a brief introduction to the system and gave simple instructions on how to use it. 
The task was printed on paper and handed to the users. 
All participants had prior experience using conversational AI; therefore, we did not provide a training session on its usage, ensuring basic familiarity with the conversational interface. 
Users were informed that they could discontinue the task at any time if they wished and were instructed to explicitly indicate when they believed they had completed the task.

To assess the impact of the proposed system, we measured participants' self-efficacy before and after the task using a task-specific self-efficacy survey. 
The survey consisted of six questions (see Table~\ref{tab:rsa_self_efficacy_survey}). 
The first and last questions assessed the participants' belief in their understanding of the concepts necessary to solve the task and their ability to explain the problem-solving process. 
Questions 2 to 5 measured their confidence in completing the processes required to achieve the subtasks necessary for solving the problem. 
All items were designed to be answered on a 5-point Likert scale (1:very unconfident to 5:very confident).

This task-specific survey provides a precise evaluation of participants' confidence levels related to the RSA encryption task. 
By focusing on concrete steps within the task, it ensures content validity by directly relating survey items to the essential skills required. 
By administering this survey before and after the experiment, we observed changes in self-efficacy and analyze differences between the two groups.

To further examine the system's impact, we measured cognitive load after the task was completed using the NASA Task Load Index (TLX) questionnaire~\cite{hart1986nasa}, a validated tool for assessing mental workload. 
Additionally, we surveyed user satisfaction for both groups after the task. 
The satisfaction survey included the following questions (see Table~\ref{tab:satisfaction_survey}).



Furthermore, to compare the efficiency of the task completion process, we measured participants' task time and interaction count, defined as the number of times users interacted with the conversational AI by asking questions.

By collecting data on self-efficacy, cognitive load, user satisfaction, task time, and interaction count, we aimed to comprehensively evaluate the impact of the CPG on participants' performances and experience. 

\subsection{Task and System Implementation}

\begin{figure}[h]
  \centering
  \includegraphics[width=0.48\textwidth]{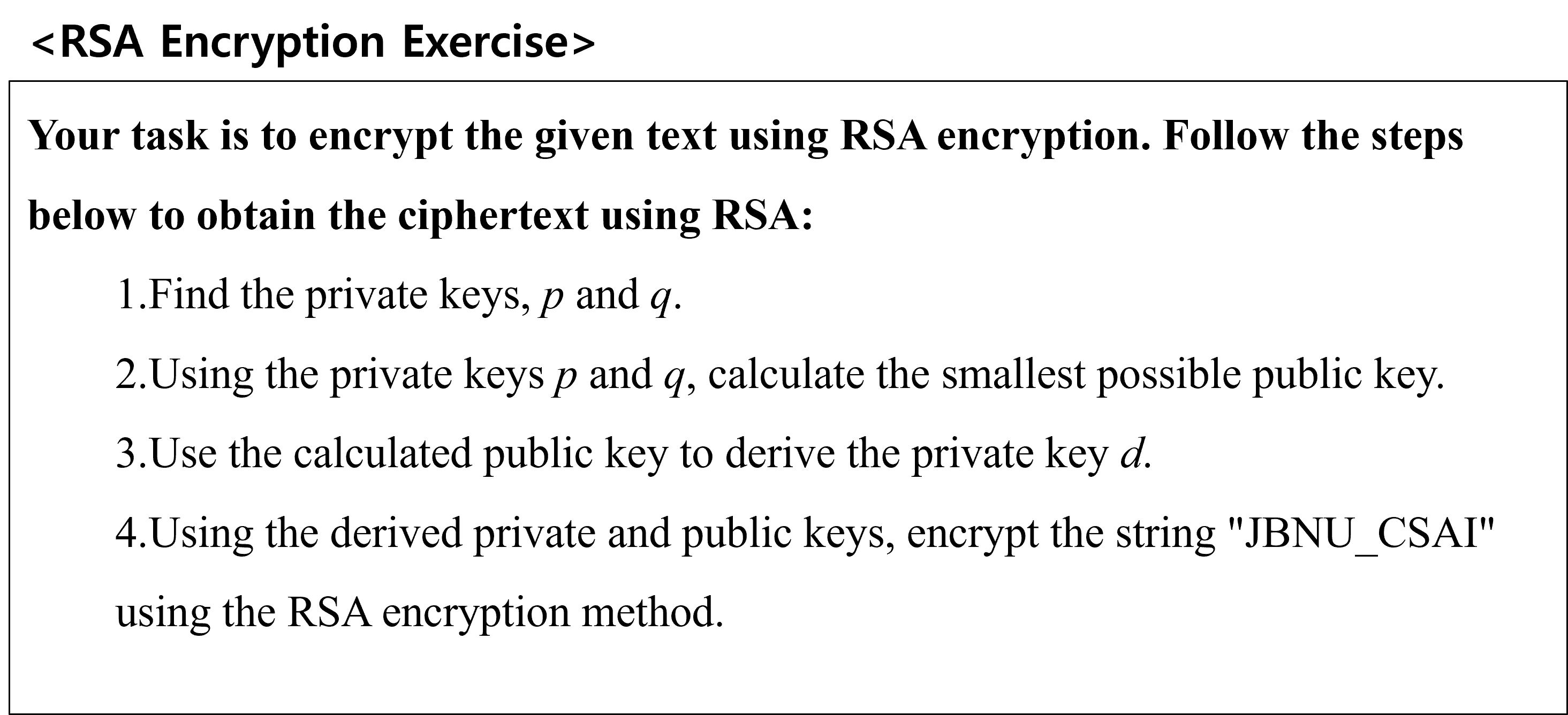}
  \caption{The task instruction for user study.}
  \Description{This figure presents the sequence of steps involved in the task for user study.}
  \label{fig:fig3}
\end{figure}

\begin{figure*}[htbp]
  \centering
  \includegraphics[width=0.64\textwidth]{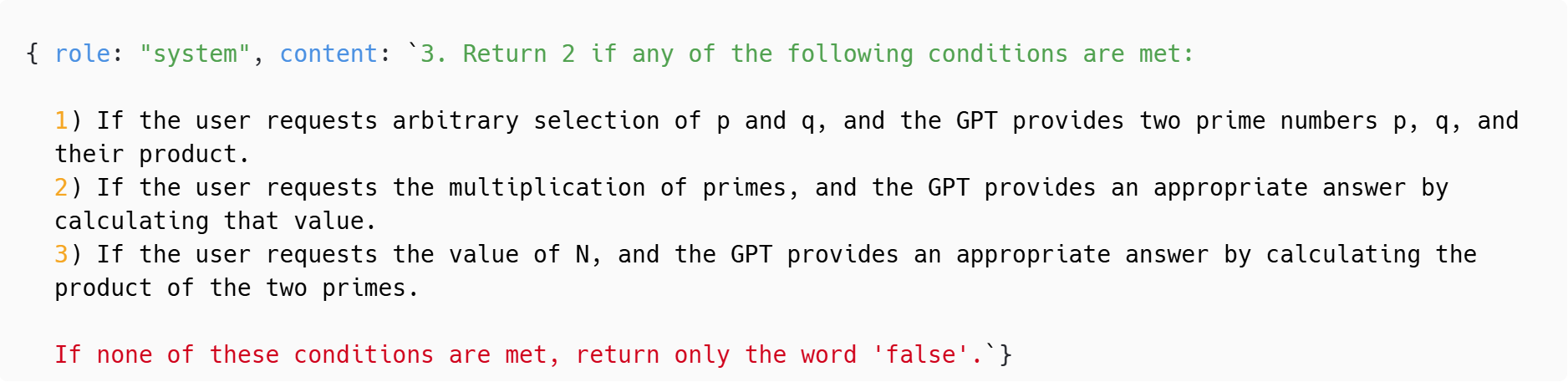}
  \caption {Evaluation rules for the “Multiplication of Primes” subtask, as used by the Progress Feedback Agent to assess completion.}
  \Description{This figure illustrates the evaluation rules for the “Multiplication of Primes” subtask, used by the Progress Feedback Agent to assess task completion and activate the corresponding progress marker.}
  \label{fig:fig4}
\end{figure*}

\paragraph{\textbf{Task}}
We designed the task as shown in Figure~\ref{fig:fig3}. 
Measuring self-efficacy through math problems is a common approach~\cite{pajares1994role,pajares1996self,huang2020better}, and encryption tasks, being inherently mathematical and involving clearly defined linear subtasks, are an appropriate choice for our experimental task.
The ultimate goal of this task is to generate an encrypted message using RSA encryption. 
Participants were provided with this final goal along with key steps associated with the subtasks required to achieve it: ``Understanding RSA Encryption Method,'' ``Multiplication of Primes,'' ``Euler's Totient Function,'' ``Public Key Generation,'' ``Private Key Generation,'' and ``Encryption Using RSA.'' For each subtasks, the Progress Feedback Agent evaluates the conversation and returns a completion value based on predefined evaluation rules. For example, the “Multiplication of Primes” step’s rules are shown in Figure~\ref{fig:fig4}.
Similar detailed rules were established for each subtask to guide the Progress Feedback Agent in its evaluations, \revt{and the comprehensive rules for the other subtasks are shown in Figure~\ref{appendix:code} (Appendix).}
Determining an appropriate number of subtasks is crucial for effective progress evaluation. 
After empirical testing, we found that 3 to 7 subtasks generally provide a manageable range.
The exact number varies by display environment.

\paragraph{\textbf{System Implementation}}
We implemented an web application based on React.js and Node.js, utilizing two GPT agents. 
For the Task Agent, we used GPT-3.5-turbo, as it has a higher likelihood of encountering errors during task execution compared to the latest LLM models. 
This choice was intentional to simulate realistic scenarios where users might face challenges, thereby allowing us to assess the impact of our system on self-efficacy under such conditions. 
For the Progress Feedback Agent, employed the GPT-4 model to more precisely evaluate the conversation topics.

When the user inputs a question into the prompt, the Task Agent generates a response. 
Additionally, the previous 10 exchanges (5 pairs of questions and answers) are stored in an array and provided to the Task Agent as part of the conversation history. 
This enables the agent to understand the context of the conversation and generate more accurate responses.

The generated question–answer pairs are then evaluated by the Progress Feedback Agent. 
The system divides the conversation goals into subtasks and stores them as objects, where each object includes attributes such as step, label, and active status, which are used to track the current progress of the conversation. 
The Progress Feedback Agent evaluates whether the conversation matches the label of each subtask, and returns the assigned value for the corresponding step if there is a match. 
The Progress Bar component observes this step value and inserts the appropriate step object into an array to visually display the user's progress through the tasks. 

The Subtask Marker is implemented to appear on the Progress Bar when its active status is set to ``TRUE.'' 
Markers are displayed in a predefined order on the Progress Bar, regardless of the sequence in which the user completes the subtasks. 
Uncompleted subtasks are not displayed on the Progress Bar, and no additional hints are provided. 
For example, if a user completes the first and third subtasks in sequence, their corresponding markers will appear consecutively on the Progress Bar. 
When the second subtask is later completed, its marker is added between the first and third markers. 
This approach is designed to clearly convey the predefined problem-solving process to users, enhancing learning outcomes.


When the Progress Feedback Agent determines that the final goal has been achieved, a modal window is presented to the user to confirm whether the task is complete or if they wish to continue the conversation. 
This approach provides users with more autonomy by allowing additional opportunities to interact further if the Progress Feedback Agent incorrectly judges the task as complete or if the user does not feel that the task has been sufficiently completed. 
When users proceed with the conversation, the system continuously tracks progress and updates the status of completed subtasks. 
If the user completes the task again, a modal window is displayed. 
This allows users to refine their work and complete the task.

\subsection{Results}

\paragraph{\textbf{Self-Efficacy}}
First, we performed a \textit{t}-test to compare the pre-task self-efficacy between the experimental group and the control group. The results showed that for all six items in the Self-Efficacy Survey, the \textit{p}-values were above 0.05, indicating no statistically significant differences between the two groups. This suggests that the two randomly assigned participants groups had similar levels of self-efficacy prior to performing the task.

\begin{figure}[h]
  \centering
  \includegraphics[width=0.48\textwidth]{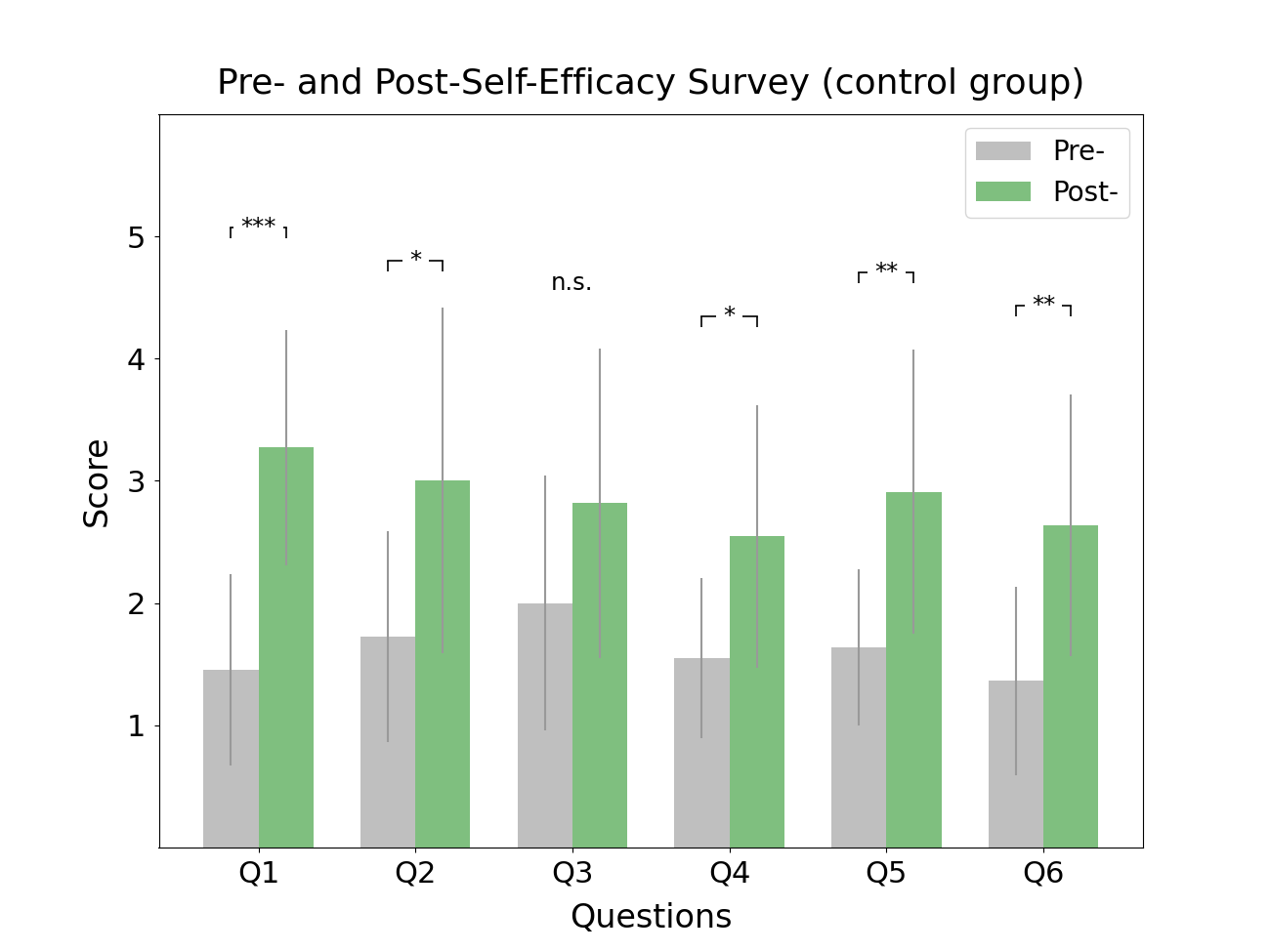}
  \caption{Average scores of Pre- and Post-Self-Efficacy Survey for the control group. Error bars represent standard deviations.}
  \Description{A bar chart displaying the average scores from the pre- and post-task self-efficacy survey for the Control Group. Each bar represents one of six survey questions, with error bars indicating standard deviation.}
  \label{fig:chart_promote(x)}
\end{figure}

We then compared the pre- and post-task self-efficacy within the control group. Figure~\ref{fig:chart_promote(x)} visualizes the scores of each item in the Self-Efficacy Survey before and after the task. Initially, we observed that the mean scores increased across all items. 
Moreover, \textit{t}-test results revealed significant differences in all items except Q3 at the 0.05 significance level, with \textit{p}-values ranging from 0.00016 to 0.02465, excluding Q3. These results suggest that performing the task using the default conversational AI Interface enhances task-specific self-efficacy.

\begin{figure}[h]
  \centering
  \includegraphics[width=0.48\textwidth]{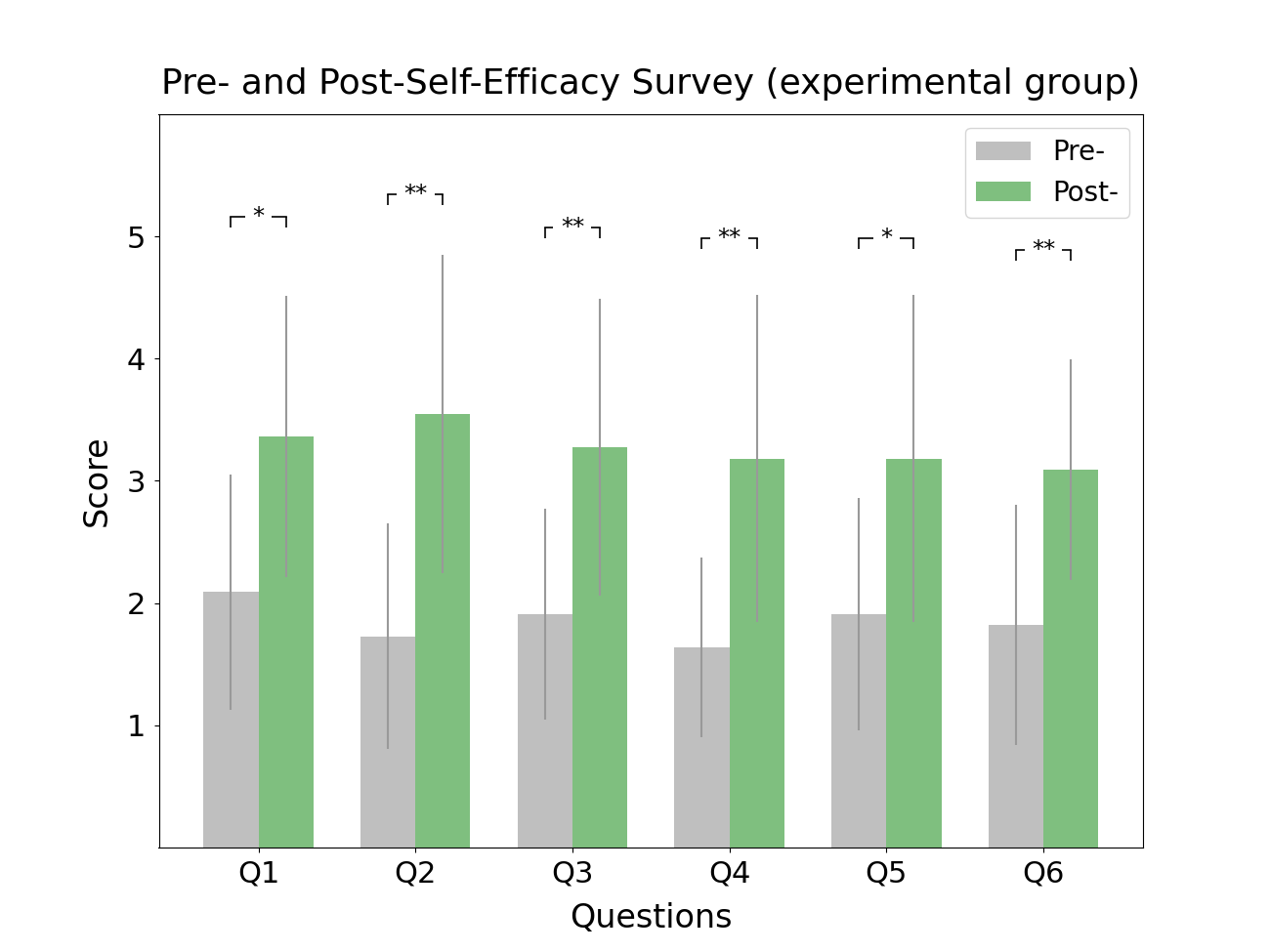}
  \caption{Average scores of Pre- and Post-Self-Efficacy Survey for the experimental group. Error bars represent standard deviations.}
  \Description{A bar chart illustrating the average pre- and post-task self-efficacy scores for the Experimental Group. The figure shows significant improvement across most questions compared to the Control Group, with error bars representing standard deviation.}
  \label{fig:chart_promote(o)}
\end{figure}

Next, we compared the pre- and post-task self-efficacy within the experimental group. Figure~\ref{fig:chart_promote(o)} visualizes the scores of each item in the Self-Efficacy Survey before and after the task. Similar to the control group, we observed increases in the mean scores across all items. 
The \textit{t}-test results revealed significant differences at the 0.05 significance level for all questions, with \textit{p}-values ranging from 0.00202 to 0.02542. These results suggest that performing the task using the CPG also enhances task-specific self-efficacy.

Q3 showed a significant improvement only in the experimental group, while all other items demonstrated significant self-efficacy improvements in both groups.
To further compare the degree of improvements of these improvements between the two groups, we calculated the effect size using Cohen's \textit{d} for pre- and post-task self-efficacy scores in each group. 
The results are presented in Table~\ref{table:effect_sizes}. 
According to these results, Q1 \rev{(experimental: 1.128 vs. control: 1.977)} and Q5 \rev{(experimental: 1.03 vs. control: 1.29)} exhibited greater improvements in the control group, suggesting a larger magnitude of change among its participants. 
However, for Q2 to Q4, the experimental group demonstrated larger improvements in self-efficacy \rev{(Q2: experimental 1.512 vs. control 1.036; Q3: experimental 1.217 vs. control 0.672; Q4: experimental 1.351 vs. control 1.07)}. 
For Q6, the extent of improvements was similar between the two groups \rev{(experimental: 1.256, control: 1.303)}.
These findings suggest that using the default conversational AI Interface also contributed to increases in self-efficacy. 
Participants in the control group also experienced improvements in self-efficacy through successful task completion and learning from problem-solving experiences. 
However, the use of the CPG facilitated greater improvements in self-efficacy across more items, further demonstrating its effectiveness in supporting task-specific self-efficacy.

\begin{table*}[ht]
\centering
\caption{Effect sizes (Cohen's $d$) for self-efficacy improvements in the experimental and control group}
\setlength{\tabcolsep}{8pt}
\renewcommand{\arraystretch}{1.3}
\begin{tabular}{lcccccc}
\hline
\toprule
 & \textbf{Q1} & \textbf{Q2} & \textbf{Q3} & \textbf{Q4} & \textbf{Q5} & \textbf{Q6} \\
\midrule
Experimental group & 1.128 & 1.512 & 1.217 & 1.351 & 1.03 & 1.256 \\ \rowcolor{gray!18}

Control group & 1.977 & 1.036 & 0.672 & 1.07 & 1.29 & 1.303 \\

\bottomrule

\end{tabular}
\label{table:effect_sizes}
\end{table*}

\begin{figure}[!ht]
  \centering
  \includegraphics[width=0.48\textwidth]{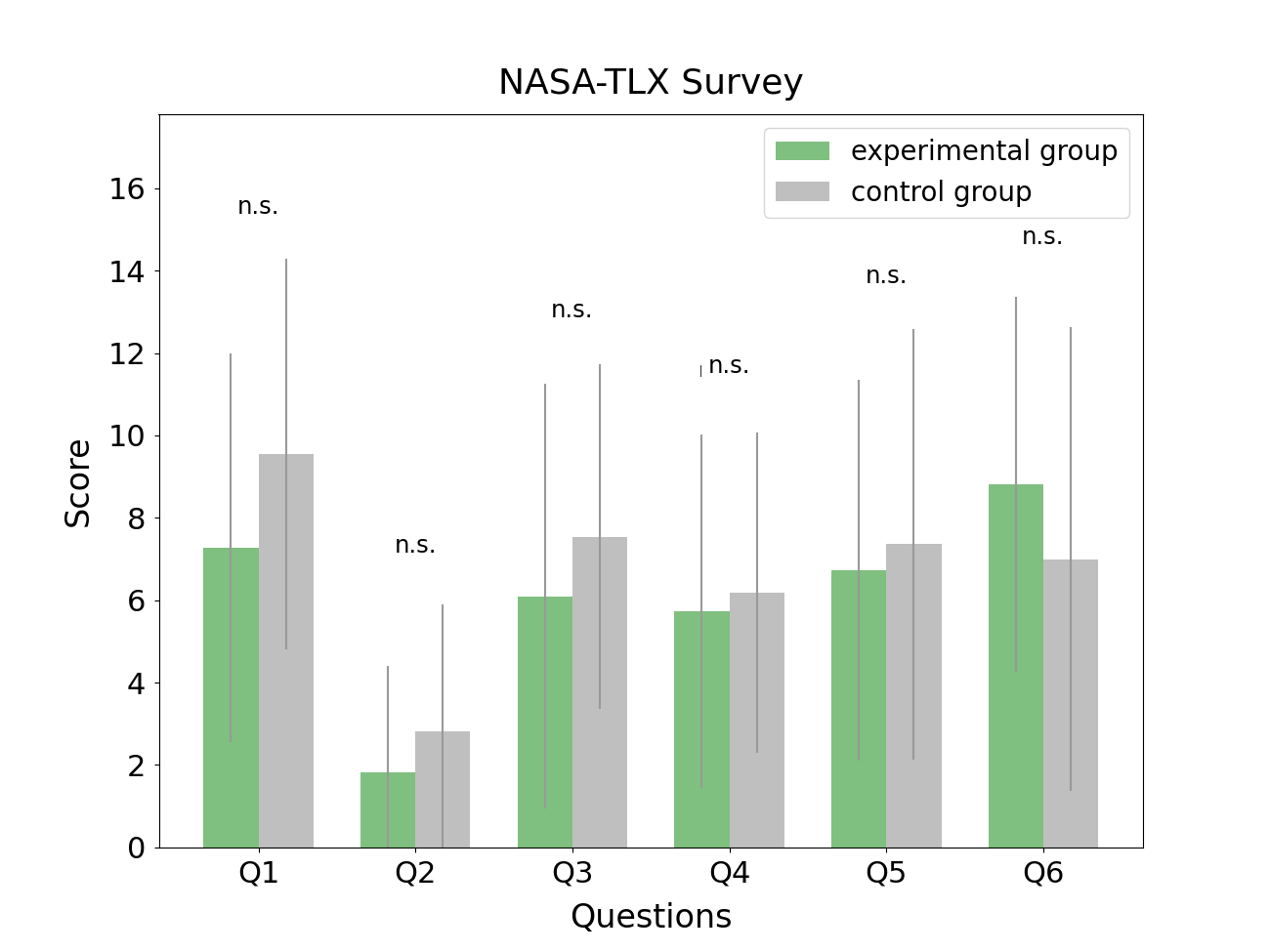}
  \caption{Average Cognitive Load scores for the experimental and control group. Error bars represent standard deviations.}
  \Description{A bar chart comparing the average cognitive load scores of the Control and Experimental groups. The chart displays no statistically significant differences between the two groups across six dimensions of cognitive load, based on NASA-TLX scores.}
  \label{fig:CL_chart}
\end{figure}

\paragraph{\textbf{Cognitive Load}}
Participants measured their cognitive load after task completion using the NASA-TLX questionnaire, and the results are illustrated in Figure~\ref{fig:CL_chart}. 
For Q1-5, the average cognitive load of the experimental group was lower than that of the control group. 
However, the \textit{t}-test results showed that the \textit{p}-values for Q1-5 were all above 0.05, indicating no statistically significant difference in the cognitive load perceived by users between the two groups. 
Additionally, Q6 also showed no significant difference, with a \textit{p}-value above 0.05. 
These findings suggest that performing the task using the CPG does not impact on user's cognitive load.




\begin{figure}[!ht]
  \centering
  \includegraphics[width=0.48\textwidth]{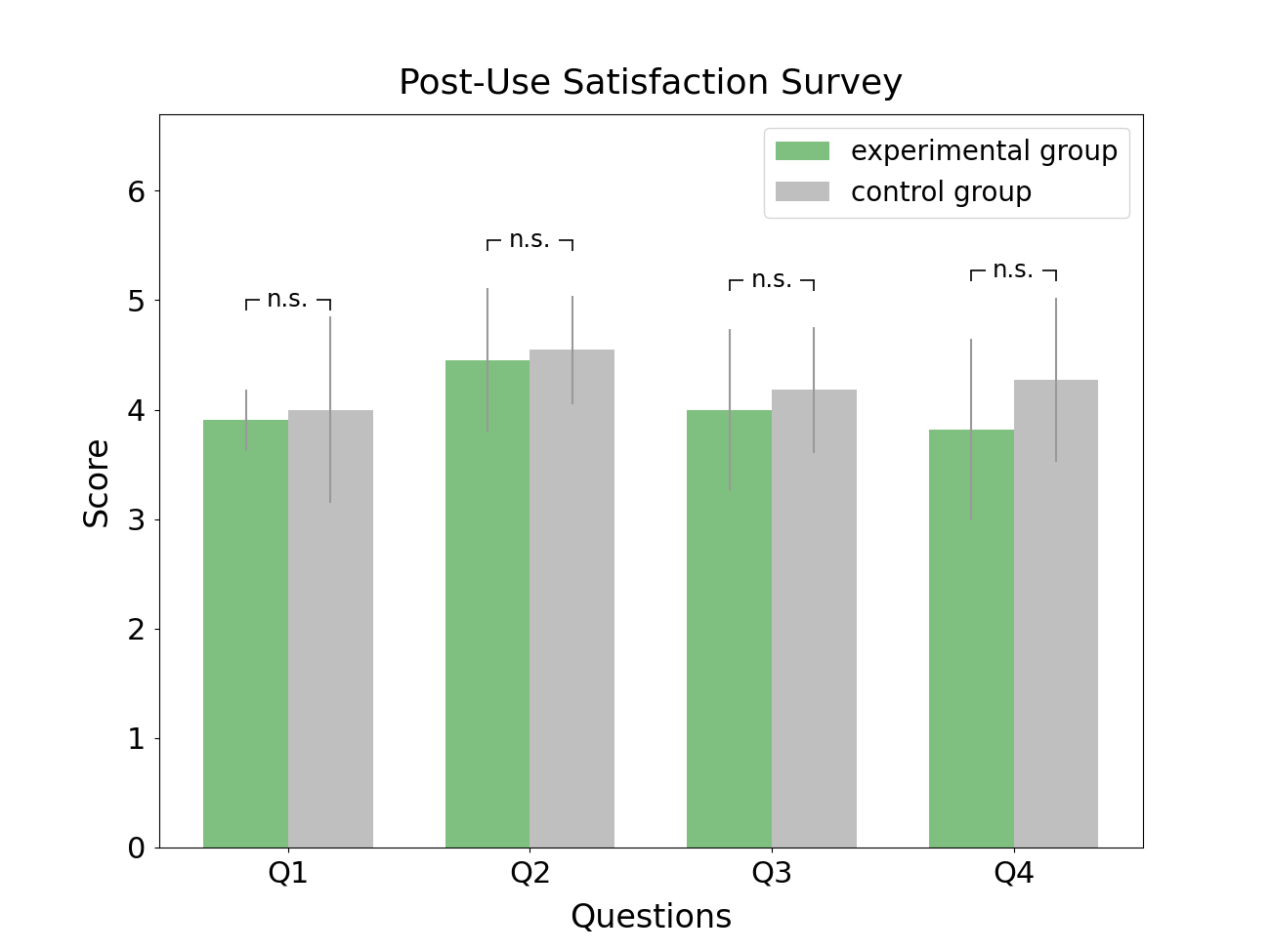}
  \caption{Average satisfaction scores for the experimental and control group. Error bars represent standard deviations.}
  \Description{This figure presents the average user satisfaction scores for both the Control and Experimental groups across four questions. No significant differences were observed between the two groups, with similar satisfaction levels reported.}
  \label{fig:saftis}
\end{figure}

\paragraph{\textbf{Satisfaction}}
Participants conducted the satisfaction survey, and the results are illustrated in Figure~\ref{fig:saftis}. 
For all questions, the difference in averages between the two groups was small, and the \textit{t}-test results showed that the p-values were all above the significance level of 0.05, indicating no statistically significant differences between the groups. 
These findings suggest that performing the task using the CPG does not impact on user's satisfaction and experience.

\paragraph{\textbf{Task Time \& Interaction Count}}
The average task time observed in the experiment was 613 seconds for the control group and 649 seconds for the experimental group. 
The \textit{t}-test results showed a \textit{p}-value of 0.897, indicating no statistically significant difference between the two groups. 
In terms of interaction count, the control group and experimental group had averages of 11.5 and 8.82, respectively. 
A \textit{t}-test resulted in a \textit{p}-value of 0.365, also showing no statistically significant difference between the groups.

The analysis revealed no statistically significant differences between the two groups in terms of average task time or the number of interactions. 
This suggests that the CPG did not lead to measurable improvements in task performance. 
Regarding task time, it can be influenced by the response time of the GPT model used in the experiment, and due to the high variability of this response time, task time may be limited as an accurate measurement metric. 

Although the difference in average task time was not statistically significant, the longest task time in the experimental group (2441 seconds) exceeded that in the control group (1458 seconds). 
This difference may explain the slightly longer average task time in the experimental group. Additionally, the reduced variance in task times suggests a more consistent user experience in the experimental group compared to the control group.
In terms of interaction count, since the steps required to perform the task were roughly outlined and users followed them similarly, the number of utterances did not differ significantly between interfaces. 
Given the current experimental settings, measuring and comparing efficiency using task time and interaction count appears challenging, indicating a need for alternative metrics or measurement methods.

\paragraph{\textbf{User Feedback}}
Below we delineate notable observations from our participants (See Table~\ref{tab:feedback}). 
Feedback about system responsiveness was reported by participants in both experimental and control groups (P2, P4), specifically noting delays in the Chat GPT API's response time, unrelated to our interface.
P14 successfully obtained the encrypted value but failed to recognize it as the correct answer, repeatedly performing the encryption process. Unable to realize task completion, P14 abandoned the experiment.
P7 completed a task in a single prompt by addressing all subtasks at once, triggering the CPG’s task completion modal. This caused confusion, as reflected in his feedback.

\begin{table*}[h]
\centering
\caption{Participants' Feedback}
\renewcommand{\arraystretch}{1.4}
\begin{tabular}{ll}
\hline
\toprule
\textbf{Group} & \textbf{Feedback} \\
\midrule
Control 
            & ``\textit{Performance improvement is necessary.}'' (P4) \\ 
            & ``\textit{I repeatedly attempted without realizing that the correct answer had already been derived.}'' (P10) \\

Experimental 
             &``\textit{The response time seems to be slow.}'' (P2)   \\
             & ``\textit{I noticed the window indicating task completion a few times, but I am not sure what triggered it.}'' (P7) \\
\bottomrule
\end{tabular}
\label{tab:feedback}
\end{table*}
\section{Discussion and Limitation}

Below we discuss the implication of results, discuss important points in detail, as well as limitations behind our work. 

\subsection{Implication of Results}

In our study, we evaluated the impact of the CPG on self-efficacy and task performance. 
Participants in both the control and experimental group showed improvements in self-efficacy. 
However, the self-efficacy of participants in the experimental group, who used the CPG, improved significantly compared to the control group. 
This suggests that the proposed system is effective in enhancing users’ self-efficacy. 
Furthermore, we observed promising indications of the benefits of CPG, such as an effective guidance tool.

Additionally, no significant differences were observed in task time, interaction count, cognitive load, or user satisfaction between the two groups. 
This indicates that the CPG does not impose additional cognitive burden or negatively affect task performance and user experience. 
Below we further discuss in detail, about the implications of our work. 

\subsection{Discussion}
We present additional discussions about our work.
\paragraph{Applicability}

While the CPG framework provides meaningful progress feedback for goal-oriented tasks that can be decomposed into linear subtasks, its applicability is limited in other situations. For instance, non-task-oriented dialogues that lack a specific goal are not well-suited to progress evaluations~\cite{hussain2019survey}. Similarly, when tasks cannot be broken down into well-defined subtasks (e.g., creative writing), the CPG can only offer final outcome feedback rather than incremental guidance. 
Moreover, although some flexibility is allowed in how subtasks are approached, the CPG remains unsuitable for open-ended or unstructured conversations.

Even with well-defined subtasks, the CPG is limited in supporting users who do not follow a linear, step-by-step approach, as observed in our user study (e.g., P7), where tasks were not completed in a structured manner as expected.
Potential solutions include integrating automated prompt suggestions to better guide users through the subtasks~\cite{su2023prompt} or providing a post-task summary of all completed steps to reinforce learning and comprehension~\cite{leopold2013learning}. We leave these issues as future work.

\paragraph{Cognitive load.} Cognitive load impacts not only user experience and task performance but also self-efficacy~\cite{carroll1997human}. 
Excessive cognitive load caused by additional interface elements can overwhelm users' cognitive resources.
This can potentially lead to decreased self-efficacy, frustration, and reduced confidence in completing tasks~\cite{sweller1988cognitive, paas2003cognitive}. 
Conversely, insufficient cognitive load does not support improvements in self-efficacy. 
An optimal level of cognitive load is crucial for maintaining user engagement and facilitating effective performance~\cite{van2005cognitive}. 
We plan to explore this issue.

\paragraph{Negative feedback in progress evaluation.} Currently, the CPG evaluates progress based on subtask completion. 
Incorporating failures as part of progress could provide users with a more comprehensive view of their performance.
Nonetheless, incorporating negative feedback requires careful consideration, as it may conflict with the provision of mastery experiences intended to enhance self-efficacy. This is especially crucial in cases of repeated failures, where negative feedback could diminish motivation. Conveying negative feedback positively, such as through humor, could mitigate these issues. For example, Dybala et al.~\cite{dybala2009activating} suggest that humor can effectively deliver negative feedback without reducing user engagement, which could be a valuable consideration for future enhancements of the CPG. 

\rev{
\paragraph{Self-efficacy measurement.} Self-efficacy is inherently linked to beliefs in specific domains, and its accurate assessment requires the refinement of measurement scales. In this study, self-efficacy was evaluated for both the control and experimental groups using a consistent measurement procedure and survey questions. However, the evaluation employed was specifically developed for the tasks used in our experiments; as Bandura~\cite{bandura2006guide} suggests, task-specific self-efficacy scales should be refined to capture the nuances of different domains. For different tasks, the development of domain-specific measurement scales (or assessment instruments) may be necessary, and a thorough review and subsequent refinement of the current scale could further enhance its accuracy. Moreover, cultural and environmental factors may influence self-efficacy outcomes, and these should be carefully considered in relation to the specific tasks and characteristics of the participants~\cite{bandura2006guide}.
}


\subsection{Limitations and Future Work}

There are limitations related to the generalizability of our findings due to the small and homogeneous sample used in this study. Future research should involve replicating the study with a larger and more diverse participant pool to increase the reliability of the results and explore the long-term effects on self-efficacy. Future work could involve generating evaluation rules or integrating machine learning techniques to enhance accuracy, thereby increasing the system’s robustness and effectiveness across diverse contexts. 

\rev{
In addition, integrating automated subtask decomposition and evaluation can enhance progress tracking, thereby enabling the system to manage complex tasks more effectively.
}
Furthermore, future work also may explore extending the CPG to accommodate less structured scenarios, such as entertainment-driven chatbot, where task goals are harder to define.
\\

\section{Conclusion}
\label{sec:conclusion}
In this study, we introduce the CPG, a system designed to provide a visual interface for representing progress in conversations with conversational AI. 
The proposed interface tracks the user's goal and related subtasks, offering visual feedback on progress as each subtask is completed. This enables users to receive feedback on their progress and offers mastery experiences by highlighting the completion of each subtask.

Results from the user study demonstrate that the CPG significantly improved self-efficacy compared to the default conversational AI interface. 
While additional measures such as task time, interaction count, cognitive load, and satisfaction are also evaluated, no statistically significant differences are observed. This suggests that the proposed interface did not negatively impact task efficiency or user satisfaction. 
In conclusion, the CPG is more effective at improving self-efficacy than the default conversational AI interface. 
This system is particularly beneficial for tasks involving multi-turn conversations, where repeated failures are likely, as it provides users with a better experience by enhancing self-efficacy.

\bibliographystyle{ACM-Reference-Format}
\bibliography{conversation-progress-guide}

\onecolumn

\newpage

\appendix

\rev{\section{Appendix}}

\begin{figure*}[h]
  \centering
  \includegraphics[width=0.96\linewidth]{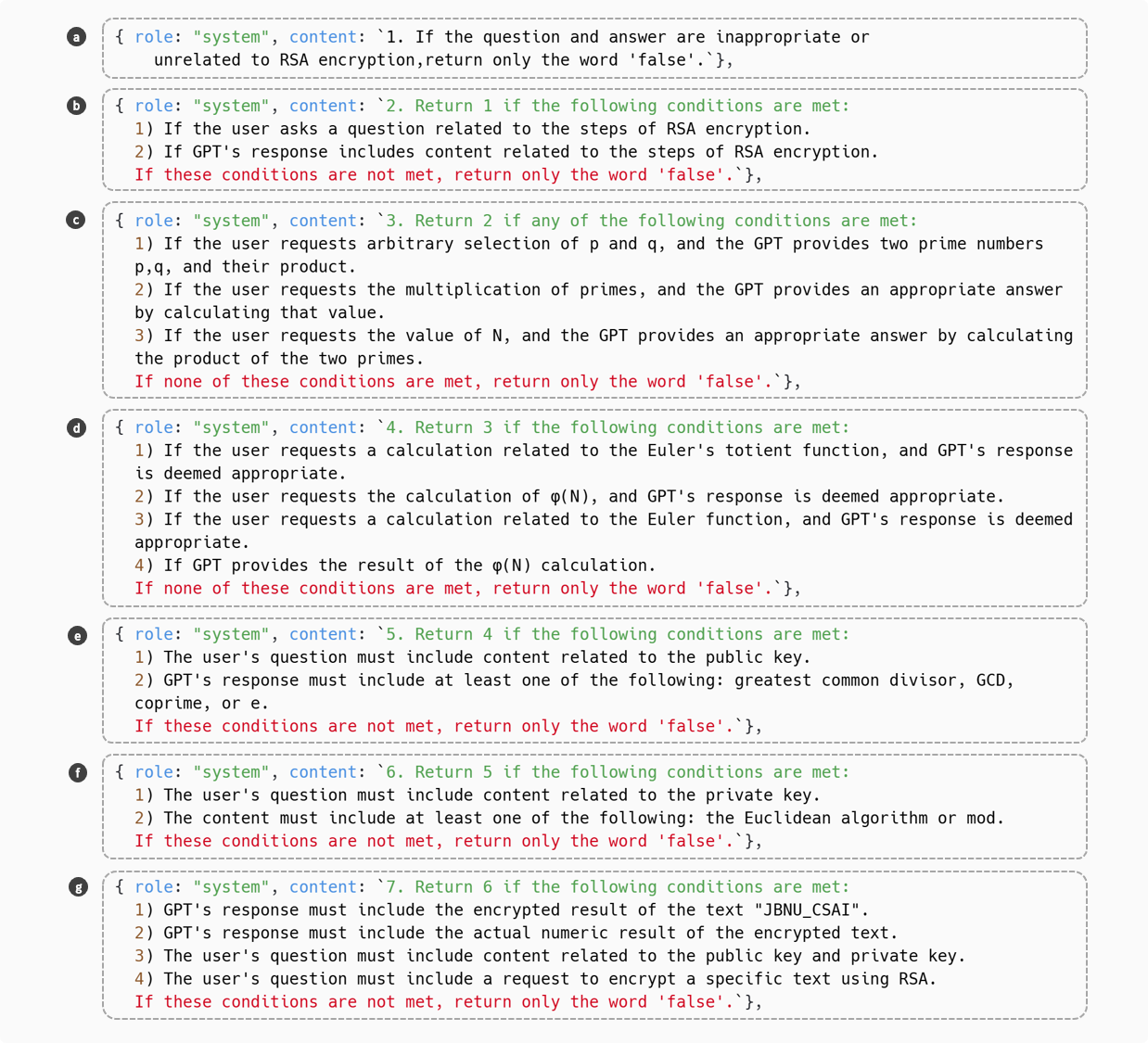}
  \caption{\rev{Evaluation rules used by the Progress Feedback Agent to assess task completion and trigger progress markers. These rules were specifically developed for this experiment and include (a) a fundamental rule applicable to all subtasks for relevance checking, as well as subtask-specific rules for: (b) Understanding RSA Encryption Method, (c) Multiplication of Primes, (d) Euler’s Totient Function, (e) Public Key Generation, (f) Private Key Generation, and (g) Encryption Using RSA.}}
  \Description{This figure illustrates the basic rules and evaluation rules for the subtasks "Understanding RSA Encryption Method", "Multiplication of Primes", "Euler’s Totient Function", "Public Key Generation", "Private Key Generation", and "Encryption Using RSA" which are used by the Progress Feedback Agent to assess task completion and activate the corresponding progress markers.}
  \label{appendix:code}
\end{figure*}


\end{document}